**ESR Probing of Quantum Critical Phenomena in Doped S=1/2 AF Quantum Spin Chain.**

S.V.Demishev[a], A.V.Semeno[a], N.E.Sluchanko[a], N.A.Samarin[a], I.E.Tarasenko[a], H.Ohta[b], S.Okubo[b]

[a]*Low Temperatures and Cryogenic Engineering Department, A.M.Prokhorov General Physics Institute of Russian Academy of Sciences, Vavilov street, 38, 119991 Moscow, Russia*

[b]*Molecular Photoscience Research Center, Kobe University, 1-1 Rokkodai, Nada, Kobe 657-8501, Japan*

Short title: ESR probing of quantum critical phenomena.

Communicating author: Professor Sergey Vasil'evich Demishev

Low Temperatures and Cryogenic Engineering Department,

A.M.Prokhorov General Physics Institute of Russian Academy of Sciences,

Vavilov street, 38

119991 Moscow, Russia

Tel./Fax: +7-495-1358129

E-mail: demis@lt.gpi.ru

**Abstract**

The results of high frequency (60-315 GHz) studies of the ESR in $CuGeO_3$ single crystals containing 0.9% of the Mn impurity are reported. The quantitative ESR line shape analysis shows that the low temperature ($T$<40 K) magnetic susceptibility of $Cu^{2+}$ chains diverges as $\chi \sim 1/T^{\alpha}$ with the critical exponent $\alpha=0.81\pm0.03$ and therefore indicates an onset of a quantum critical (QC) regime. The scenario, in which disorder caused by the Mn impurity in the quantum spin chains in $CuGeO_3$ may lead to the co-existence of the QC regime and the spin-Peierls dimerisation, is discussed. For the quantitative description of the temperature dependences of the line width and *g*-factor a model assuming the crossover from the high temperature semiclassical Nagata and Tazuke limit to the low temperature quantum case described by Oshikawa and Affleck theory is suggested.

**1. Introduction.**

Recently it has been shown [1-5] that doping of the spin-Peierls compound $CuGeO_3$ with magnetic impurities such as iron and cobalt gives rise to disorder driven quantum critical (QC) phenomena. The insertion of magnetic ions with S=3/2 (Co) or S=2 (Fe) at a concentration level $x$=1-2% into antiferromagnetic (AF) $Cu^{2+}$ quantum spin chains (S=1/2) has led to a reduction of both the spin-Peierls and Neel transitions at least down to 1.8 K or 0.5 K in the cases of Co and Fe respectively [1-5]. The ground states for $CuGeO_3$:Fe and $CuGeO_3$:Co are expected to be a Griffiths phase (GP) [1-5], for which the theory [1,6,7] predicts a power law for magnetic susceptibility,

$$\chi(T) \sim 1/T^{\alpha}, \qquad (1)$$

different from the Curie-Weiss or Bonner-Fisher laws. In Eq.(1) the critical exponent satisfies the condition $\alpha<1$ and is non-universal depending on the random field characteristics [7].

For $CuGeO_3$:M (M=Fe, Co), the power law (1) with $\alpha$=0.34 (Fe) and $\alpha$=0.9 (Co) has been observed at $T<T_G \sim 40$ K [1-5] and lasted down to the lowest temperature. Thus the experimental data [1-5] show that the transition into the GP occurs at temperatures, which considerably exceed all characteristic temperatures on the $x$-$T$ phase diagram expected in the standard scenario of doping [8].

The aim of the present work is to study the magnetic properties of $CuGeO_3$ single crystals doped with Mn magnetic impurity (S=5/2) by means of electron spin resonance (ESR) technique. The advantage of the ESR consists in possibility of a direct monitoring of the magnetic properties of $Cu^{2+}$ quantum spin chains, whereas the use of the static magnetization measurements implies an ambiguous procedure of the separation of the chains and possible paramagnetic impurities contributions. The first paper, where CuGeO$_3$:Mn system has been mentioned, is Ref. 9, but no details of physical properties of this material have been reported in this paper. A short note about our ESR experiments on CuGeO$_3$:Mn has been published in [10]. Here we will present detail report about high frequency (60-315 GHz) measurements focused on the quantum critical phenomena and ESR parameters.

**2. Experimental details.**

The ESR experiments were carried out in the temperature range 1.8-70 K using cavity spectrometer (frequency range 60-100 GHz and magnetic field *B* up to 7 T) and quasi-optical setup (frequency range 100-315 GHz and magnetic field *B* up to 12 T). In the latter case the transmission of the microwave radiation through the sample was measured as a function of magnetic field. The CuGeO$_3$ single crystals containing 0.9% of Mn impurity have been studied. The X-ray and Raman scattering technique controlled the quality of samples; the actual content of impurity in the sample was checked by chemical analysis. From the analysis of the structural data it was possible concluding that Mn impurity similarly to Fe and Co [1-4] substitute cooper in chains. The synthesis technique details can be found elsewhere [11]. In ESR measurements, the magnetic field was aligned along the crystallographic *a* axis.

**3. Experimental ESR spectra.**

The observed ESR spectrum is represented by a single Lorentzian line. No extra lines, which can be attributed to an antiferromagnetic resonance mode expected in the standard scenario of doping [8], were found in the whole temperature/frequency/magnetic field domain studied. At fixed temperature the resonance shifts almost linearly with frequency in the range 60-315 GHz. As long as the ESR line is relatively broad the best spectra were obtained for frequencies exceeding 200 GHz (fig. 1,2). It is visible from fig. 1-2 that this line broadens and shifts with a lowering temperature and for this frequency range all spectroscopic parameters, line width, $g$-factor and integrated intensity were obtained as functions of temperature. Although the ESR line can be attributed to collective ESR on $Cu^{2+}$ chains modified by doping with $Mn^{2+}$ ions the $g$-factor value for $T > 70$ K is only $g \sim 2.10$ and considerably lower than the $g$-factor $g = 2.15$ for the undoped $CuGeO_3$ in the case ***B***ǁ***a*** [1-5]. Below we consider the possible reason for this discrepancy, as well as the temperature dependence of the integrated intensity and other parameters of the ESR.

**4. Discussion.**

4.1 *Integrated intensity and quantum critical behavior.*

Basing on the analogy with the previous results obtained for $CuGeO_3$ doped with Fe and Co impurities [1-5] and on the proportionality of the integrated intensity $I$ to magnetic susceptibility, $I(T) \sim \chi(T)$, it is possible to expect a power law (1) describing temperature dependence $I(T)$. As follows from fig. 3, the asymptotic power law (1) with $\alpha = 0.81 \pm 0.03$, which is a fingerprint for the disorder driven QC regime, begins at $T_G \sim 40$ K. This is in agreement with the temperatures observed for the transition into the Griffiths phase in $CuGeO_3$:Fe and $CuGeO_3$:Co [1-5]. In the vicinity of the charactrestic temperature $T_D = 16$ K, the integrated intensity starts to deviate from the power law (fig. 3). The value of $T_D$ is close to the spin-Peierls transition temperature $T_{SP} = 14.5$ K in pure $CuGeO_3$ but somewhat higher.

Interesting that below $T \sim 7$ K for the frequency 210 GHz and below $T \sim 5$ K for the frequency 210 GHz a power law with the same $\alpha$ is restored and holds up to the lowest temperature studied (fig. 3). The observed $I(T)$ dependence is different from that reported for $CuGeO_3$:Fe and $CuGeO_3$:Co [1-5] and thus reflects a characteristic effect of the Mn impurity.

Such an unusual behavior may be qualitatively explained as follows. As long as at $T_D$ the integrated intensity (and hence $\chi(T)$) starts to decrease this characteristic temperature may be attributed to dimerisation of the quantum spin chains. A condition $T_D<T_G$ means that the possible spin-Peierls transition in CuGeO$_3$:Mn occurs within the Griffiths phase, i.e., when the magnetic subsystem is divided into spin clusters with the different coupling constants [6-7]. As long as the temperature lowering leads to an increase of the cluster size [6-7], a transition into the dimerised state may be allowed for the chains belonging to clusters with sizes exceeding the coherence length. Below $T_D$, the magnetic contribution from dimerised Cu$^{2+}$ chains vanishes rapidly due to the opening of a spin gap, and only the chains in the QC state contribute to the susceptibility. Therefore, a power law for $\chi(T)$ and $I(T)$ may be restored at $T<T_D$. As long as in the disorder driven QC regime the index α only depends on the space dimension and dynamic exponent connecting time and length scales [12], it is possible to expect a re-entrant quantum criticality at $T<T_D$ with the same value of the critical exponent as observed experimentally (fig. 3). Moreover the fact, that in our case the condition $T_D>T_{SP}$ is fulfilled, may be a consequence of the change of the phonon-magnon interaction in a finite spin clusters.

4.2 *Line width and g-factor.*

The experimental spectra suggest that the *g*-factor for CuGeO$_3$:Mn vary with temperature in a non-monotonous way (fig. 1-2). For $T<70$ K the *g*-factor first decreases and than, below $T\sim10$ K, starts to increase (fig. 4). It is worth noting that the magnitude of the low temperature *g*-factor growth is more pronounced for the frequency 210 GHz and considerably damped for the frequency 315 GHz (fig. 4).

The observed $g(T)$ temperature dependence is different from the cases of CuGeO$_3$:Fe and CuGeO$_3$:Co where *g*-factor increase with lowering temperature [1-5]. The quantitative analysis of the ESR spectra [13,14] have shown that this effect may be explained in the Oshikawa and Affleck (OA) theory for ESR in 1D S=1/2 AF quantum spin chain [15,16] assuming influence of the staggered field (SF). A possible mechanism of the appearance of the staggered component of magnetization in $CuGeO_3$ doped with magnetic impurities was considered in [17].

Therefore the data in fig. 4 suggest that for $CuGeO_3$:Mn the *g*-factor may be represented as a sum of two temperature dependent parts

$$g(T)=g_0+\Delta g_{SF}(T)+\Delta g_1(T), \qquad (2)$$

where $\Delta g_{SF}$ denotes contribution from the staggered field, $\Delta g_1$ stands for the term responsible for the *g*-factor increase with temperature and $g_0$ denotes the "bare" *g*-factor value without any corrections. Taking into account the character of the expected $\Delta g_1(T)$ temperature dependence and the fact that in our case magnetic field is perpendicular to the chain direction, it is possible attempting to use Nagata and Tazuke mean field theory [18,19]. If the spin correlations are not so strong, it is possible to express $\Delta g_1(T)$ as [18-21]

$$\Delta g_1(T)=a \cdot M(T)/M_{sat}, \qquad (3)$$

where $M(T)$ and $M_{sat}$ are the magnetization for the resonant field and saturation value respectively and *a* denotes a numerical coefficient.

In the frame of this assumption the line width *w* should be also considered as a superposition of two terms

$$w(T)=w_{SF}(T)+w_1(T) \qquad (4)$$

having the same physical origin as in the case of $g$-factor (Eq. (2)). Theories predict that both $w_{SF}$ and $w_I$ should increase when temperature is lowered [15,16,19].

The experimental data for the line width in $CuGeO_3$:Mn are shown in fig. 5. In agreement with the theoretical expectations this parameter increases with lowering temperature. The increase of frequency also leads to the broadening of the ESR line; however, for the frequency 315 GHz the section of the low temperature growth of the line width at $T<10$ K is damped. As long as this temperature range is likely controlled by the staggered field effects the simultaneous decrease of the $g$-factor and line width magnitude in low temperature region for 315 GHz should correspond to a field-induced damping of the staggered component in magnetization. This effect was in detail investigated in [14] for the case of $CuGeO_3$:Fe and can be qualitatively explained by the competition between the staggered Zeeman energy and antiferromagnetic interactions in $CuGeO_3$ doped with magnetic impurities.

For the frequency 210 GHz, when the staggered field contribution could not be neglected, we attempted to check a described above model quantitatively. In the OA theory the line width in the presence of the staggered field should diverge as $w_{SF} \sim 1/T^2$ [15,16]. At the same time the temperature dependence of the staggered magnetization, which is supposed to be important in doped $CuGeO_3$ [14,17], may change the temperature asymptotic of the line width to $w_{SF} \sim 1/T$ [17]. For that reason we have chosen the model form $w_{SF}=A/T^n$ (where $n=1,2$) for the staggered field contribution in Eq.(4). For the simplicity of calculations the similar expression for $w_I=b/T^\beta$ instead of the exact one [19] has been used.

It if found that the experimental data in fig. 5 could be fitted for $n=1$ only, which is in agreement with the expected temperature dependence of the staggered magnetization caused by the competition with antiferromagnetic interactions [14,17]. The solid line in fig. 5 shows the result of the three-parameter fit well describing the shape of the experimental curve with the dominating low temperature contribution from the $w_{SF}$.

A characteristic feature of the OA theory is the universal relation between $w_{SF}$ and $\Delta g_{SF}$ [13]

$$\frac{w_{SF}}{\Delta g_{SF}} = 1.99 \frac{k_B T}{\mu_B}, \quad (5)$$

which does not depend on the staggered field magnitude. The $w_{SF}(T)$ obtained from the fitting of the data in fig. 5 have been used to calculate $\Delta g_{SF}(T)$ in Eq.(2) and thus this quantity has no free parameters. Following Eq.(3) a simple model form $\Delta g_1(T)=a \cdot \tanh(C/T)$ was chosen, and consequently another three-parameter fit was required to reproduce a complicated $g(T)$ dependence. The outcome of the aforementioned procedure is given in fig. 4 by the solid line; the good representation of the experimental data is visible. Interesting that the fit have provided value $g(T\rightarrow\infty)=2.14\pm0.02$, which is in agreement with the $g$-factor $g=2.15$ in undoped $CuGeO_3$ for the studied experimental geometry **B**||**a**.

4.3 *Conclusion.*

In conclusion, we show that the magnetic properties of $Cu^{2+}$ quantum spin chains in $CuGeO_3$ doped by Mn are qualitatively different from the properties observed for the Fe and Co doped chains. An unusual co-existence of the QC regime described by the universal critical exponent $\alpha=0.81\pm0.03$ and spin-Peierls dimerisation is suggested. The checking of this scenario, as well as additional verification of the possible size effect on the dimerisation temperature requires more theoretical and experimental studies.

The suggested approach, assuming the crossover from the high temperature semiclassical Nagata and Tazuke limit to the low temperature quantum case described by Oshikawa and Affleck theory, allows simultaneous description of the line width and *g*-factor in the frequency range ~200 GHz. The deviations from this model at higher frequencies in $CuGeO_3$:Mn may be explained by an enhancement of antiferromagnetic interactions in high magnetic fields. Further examination of this model requires ESR measurements with the magnetic field aligned along chains direction, which are considered as a task for the future investigations.

**Acknowledgements.**

The support from the RFBR grant No. 04-02-16574 and RAS Programme "Strongly correlated electrons" is acknowledged. This work was partly supported by Grant-in-Aid for Scientific Research (B) 16340106 from the Japan Society for the Promotion of Science (JSPS).

**Figure legends.**

Fig. 1. ESR spectra obtained for 210 GHz using a quasi-optical technique.

Fig. 2. ESR spectra obtained for 315 GHz using a quasi-optical technique.

Fig. 3. Temperature dependences of the integrated intensity for 210 GHz and 315 GHz. Dashed lines represent a power asymptotic behavior of the magnetic susceptibility (Eq. (1)). Arrows mark the transition into the Griffiths phase ($T_G$) and dimerisation temperature ($T_D$).

Fig. 4. Temperature dependences of the *g*-factor for 210 GHz and 315 GHz. Solid line corresponds to the model calculation (see text); dashed line is a guide to the eye.

Fig. 5. Temperature dependences of the line width for 210 GHz and 315 GHz. Solid line corresponds to the model calculation (see text); dashed line is a guide to the eye.

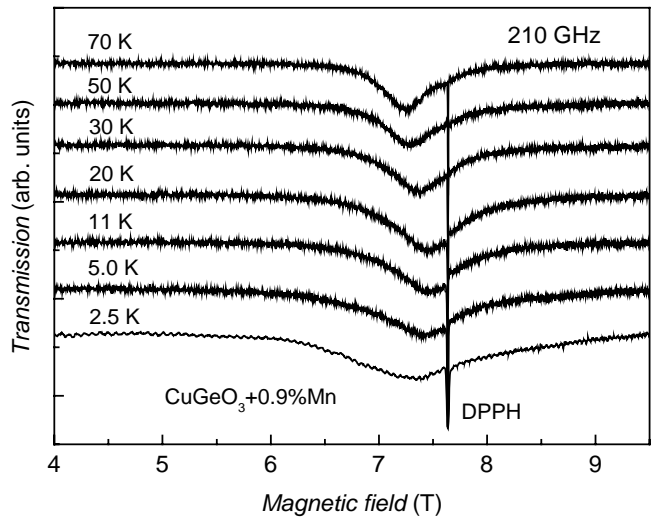

Figure 1. *S.V.Demishev et al.*

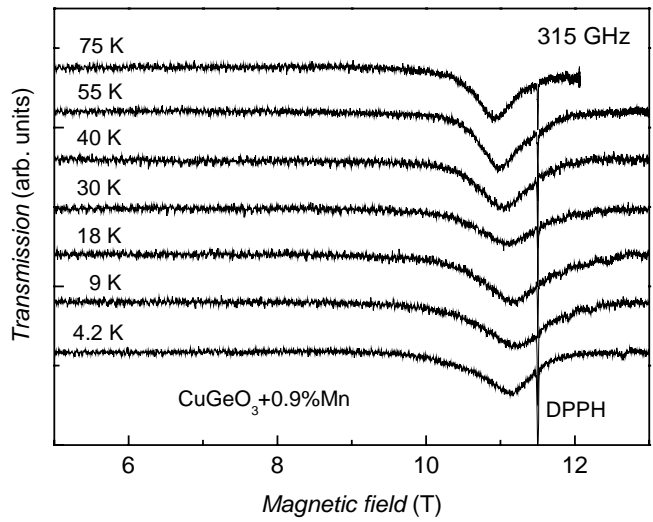

Figure 2. *S.V.Demishev et al.*

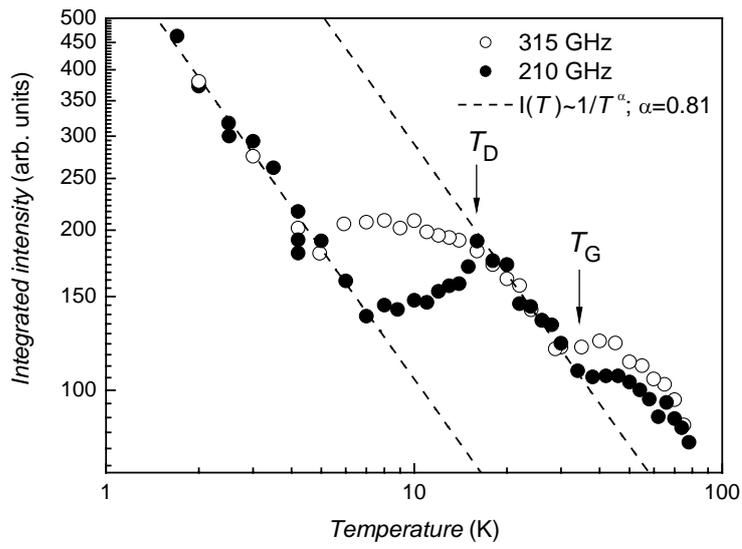

Figure 3. *S.V.Demishev et al.*

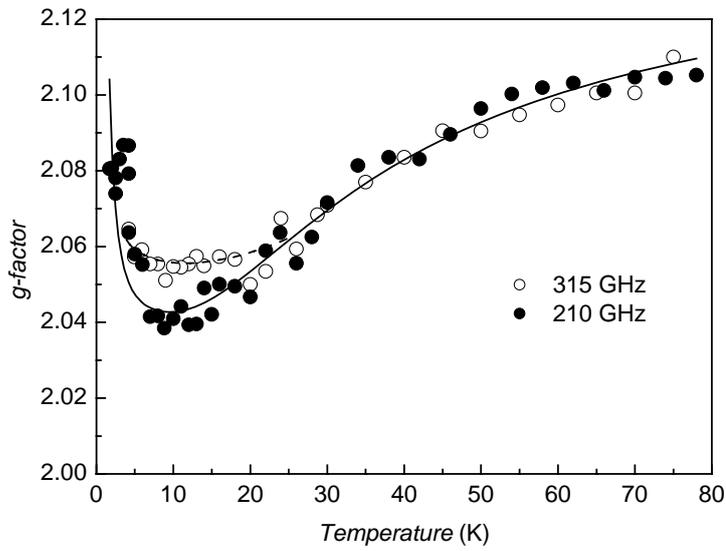

Figure 4. *S.V.Demishev et al.*

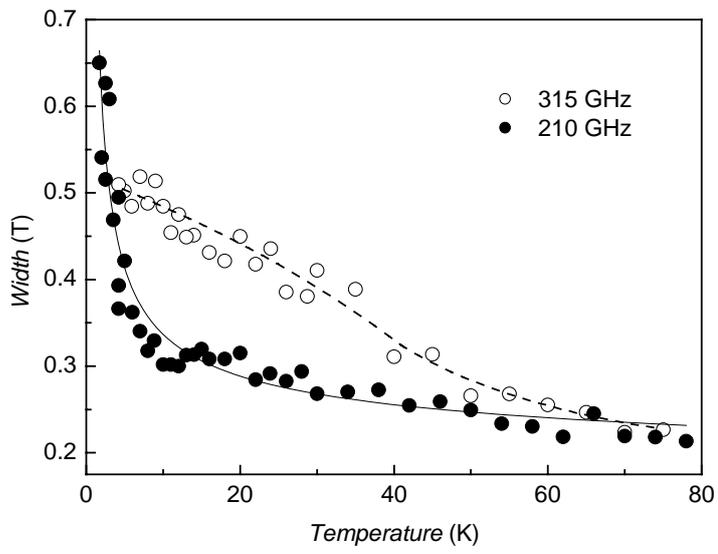

Figure 5. *S.V.Demishev et al.*